\documentclass[conference]{IEEEtran}
\IEEEoverridecommandlockouts
\usepackage{cite}
\usepackage{amsmath,amssymb,amsfonts}
\usepackage{algorithmic}
\usepackage{graphicx}
\usepackage{textcomp}
\usepackage{xcolor}
\usepackage{subfigure}
\usepackage{booktabs}
\usepackage{multirow}
\usepackage{threeparttable}
\usepackage{makecell}
\usepackage{balance}
\usepackage[ruled, linesnumbered, vlined]{algorithm2e}
\usepackage{algorithmic}

\def\BibTeX{{\rm B\kern-.05em{\sc i\kern-.025em b}\kern-.08em
    T\kern-.1667em\lower.7ex\hbox{E}\kern-.125emX}}

\begin{document}

\title{Deadline and Priority Constrained Immersive Video Streaming Transmission Scheduling}

\author{
\IEEEauthorblockN{Tongtong Feng}
\IEEEauthorblockA{\textit{Department of}\\
\textit{Computer Science and Technology} \\
\textit{Tsinghua University}\\
Beijing, China \\
fengtongtong@tsinghua.edu.cn}
\and
\IEEEauthorblockN{Qi Qi}
\IEEEauthorblockA{\textit{SKL-NST} \\
\textit{Beijing University of Posts}\\
\textit{and Telecommunications}\\
Beijing, China \\
qiqi8266@bupt.edu.cn}
\and
\IEEEauthorblockN{Bo He}
\IEEEauthorblockA{\textit{SKL-NST} \\
\textit{Beijing University of Posts}\\
\textit{and Telecommunications}\\
Beijing, China \\
hebo@bupt.edu.cn}
\and
\IEEEauthorblockN{Jingyu Wang$^{\dag}$\thanks{$^{\dag}$Corresponding author. SKL-NST is the State Key Laboratory of Networking and Switching Technology.}}
\IEEEauthorblockA{\textit{SKL-NST} \\
\textit{Beijing University of Posts}\\
\textit{and Telecommunications}\\
Beijing, China \\
wangjingyu@bupt.edu.cn}
}

\maketitle

\begin{abstract}
Deadline-aware transmission scheduling in immersive video streaming is crucial. The objective is to guarantee that at least a certain block in multi-links is fully delivered within their deadlines, which is referred to as delivery ratio. Compared with existing models that focus on maximizing throughput and ultra-low latency, which makes bandwidth resource allocation and user satisfaction locally optimized, immersive video streaming needs to guarantee more high-priority block delivery within personalized deadlines. In this paper, we propose a deadline and priority-constrained immersive video streaming transmission scheduling scheme. It builds an accurate bandwidth prediction model that can sensitively assist scheduling decisions. It divides video streaming into various media elements and performs scheduling based on the user's personalized latency sensitivity thresholds and the media element's priority. We evaluate our scheme via trace-driven simulations. Compared with existing models, the results further demonstrate the superiority of our scheme with 12{\%}-31{\%} gains in quality of experience (QoE).
\end{abstract}


\section{Introduction}
Immersive video streaming applications, such as Virtual Reality gaming (VR)\cite{VR}, 360 live streaming\cite{360}, and multiview cloud gaming\cite{AR}, have been experiencing dramatic growth over the past decade, attracting millions of active users. HTTP adaptive streaming\cite{VHAS} (HAS) also has become the common transmission framework to offer viewers an excellent quality of experience (QoE).

Deadline-aware transmission scheduling in immersive video streaming is crucial. Traditional scheduling models \cite{S, S1, S2} only work for pursuing maximizing throughput and ultra-low latency, which makes bandwidth resource allocation and user satisfaction locally optimized. In HAS\cite{VHAS}, the video streaming is divided into blocks that are encoded at different representations and offered as HTTP objects to the streaming clients. Clients only can decode blocks for playback when the entire block is received. That means each block has to be fully received within a strict deadline, otherwise it is of no use. So some deadline-constrained scheduling models\cite{DS, DS2, RS1, frame} are proposed, guaranteeing that at least a block is fully delivered within their deadlines, which is referred to as {\it delivery ratio}.

Those models can be grouped into three categories. Block-based models\cite{BS, BS1} first schedule blocks with the smallest proportion of remaining packets. They assume that all packets arrive at the beginning of blocks and are scheduled before the end of blocks. Greedy-based models\cite{GS,GS1} are based on the largest-deficit-first (LDF) policy, which is based on the longest-queue-first (LQF) policy and the smallest-remaining-proportion-first policy. Buffer-based models\cite{RS, RS2} assign each block weight and deadline, and schedule to maximize the total weight of transmitted blocks for the input sequence. However, they are not suitable for high-throughput immersive video streaming applications. 1) They can trigger frequent packet loss and re-transmissions when the deadline is variable. 2) The operation that all packets arrive at the beginning of blocks greatly increases the latency overhead. 3) {\it The optimal QoS is not equal to the optimal QoE}, in immersive video streaming, guaranteeing user requirements is more important.

Users have personalized deadlines and priority requirements for different video applications and media elements\cite{PR, PR1}. Such as VR has a strict motion-to-photon deadline (25ms) to avoid motion sickness; however, video conferencing only has a 100ms{'}s deadline on average. 360 video has gradually increasing latency requirements for control signaling (15ms), audio signaling (400ms), and video signaling (800ms). In the immersive battle game, the priority of control signaling is higher than audio signaling and video signaling. Therefore, how constructing a scheduling scheme according to personalized user requirements is a great challenge.

\begin{figure}[t]
	\centering
	\includegraphics[width=\linewidth]{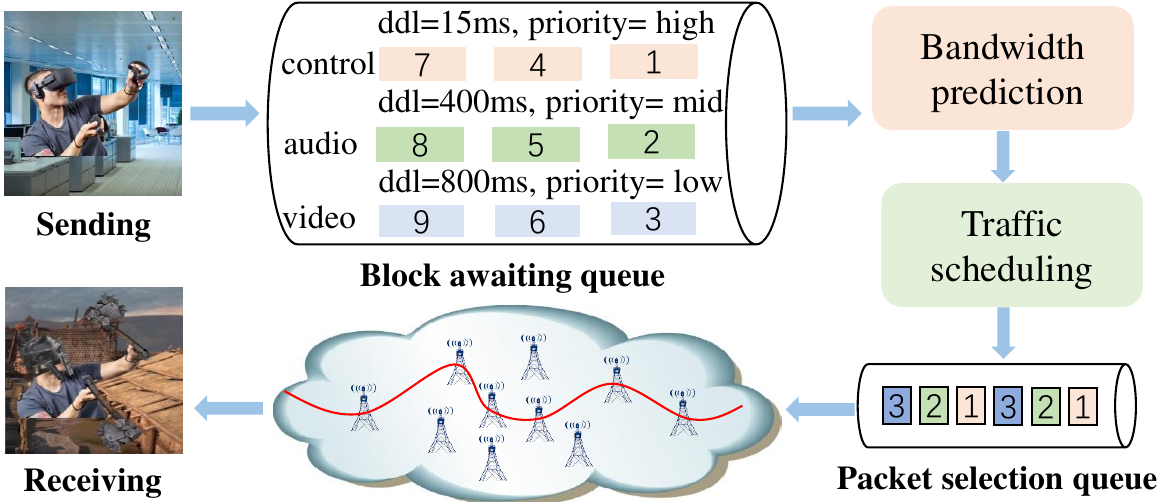}
	\caption{The overview of deadline and priority constrained immersive video streaming transmission scheduling scheme.}
	\label{struct}
\end{figure}

In this paper, we start with a close investigation of the users' personalized requirements and propose a deadline and priority-constrained immersive video streaming transmission scheme (see in Fig. \ref{struct}). It can make intelligent decisions for the sender based on the massive amount of real-time information from the network conditions and user requirements to accommodate accurate scheduling decisions with maximum user satisfaction. We first build an accurate bandwidth prediction model that can sensitively assist scheduling decisions, prefetching larger blocks or selecting high-priority blocks. We then divide video streaming into various media elements and perform scheduling based on the user's personalized deadline (latency sensitivity thresholds) and the media element's priority. In high network bandwidth, we can preload the larger media elements in the media list instead of transmitting them in order, which can meet more media elements' deadline requirements. On the contrary, we can send higher-priority media elements to alleviate the user experience's drastic fluctuation caused by network bandwidth oscillations.

We evaluate our scheme via trace-driven simulations. Compared with existing models, the results further demonstrate the superiority of our scheme with 12{\%}-31{\%} gains in QoE. Meanwhile, our scheme has an excellent generalization for various network conditions and video scenarios.

\section{Scheme design}
\subsection{Problem Definition}
We first divide video streaming as $K$ media elements $\mathbb{K}$, where $K = |\mathbb{K}|$, such as control signaling, audio signaling, and video signaling. We then divide each media element $k$ into blocks $k_v$ and save them in the block awaiting queue, where $V = |{k_v}|$. Each block $k_v$ contains many packets and has two attributes $[D_{k,v}, P_{k,v}]$, where $D_{k,v}$ represents the block $k_v$'s deadline requirement (latency sensitivity thresholds) for the current user $u$ and $P_{k,v}$ represents the block $k_v$'s priority requirement for the current user $u$. In each time slot $t$, traffic scheduling models can select one block and push the next packet of this block into the packet selection queue. When the sender receives a lost signal, the traffic scheduling model makes the lost packet into the packet selection queue. 

We consider two QoE metrics\cite{RS2} for each viewer $u$: delivery ratio $R_{k,v}$ and channel utilization $U_{k,v}$. We assume that $E_{k,v}$ is the time when the block $k_v$ arrives at the server and $F_{k,v}$ is the time when the block $k_v$ is fully-received time by the client. $R_{k,v} = 1$ (resp. 0) indicates that the block $k_v$ (resp. not) meets the user{'}s deadline requirement (delivery ratio).
\begin{equation}
R_{k,v} =\left\{
\begin{array}{rcl}
1 & & {F_{k,v} - E_{k,v} \leq D_{k,v}}\\
0     &      & {F_{k,v} - E_{k,v} > D_{k,v}}\\
\end{array} \right. 
\end{equation}

$U_{k,v} = R_t / C_t$ indicates current network bandwidth utilization, where $R_t$ indicates client's real-time received data size and $C_t$ is the known link capacity. We can calculate the QoE of viewer $u$ as follows:
\begin{equation}
QoE	= R_{k,v} + \alpha U_{k,v}
\end{equation}
where $\alpha$ is the weighted parameters to tune the channel utilization penalty. In immersive video streaming, the key problem is how to construct a scheduling scheme according to personalized user requirements to achieve maximum QoE.

\begin{figure}[b]
	\centering
	\includegraphics[width=0.9\linewidth]{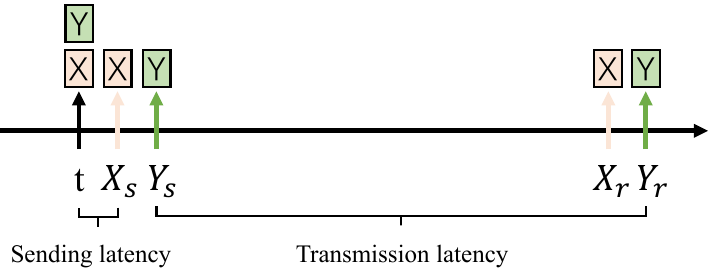}
	\caption{Bandwidth prediction model.}
	\label{BPM}
\end{figure}

\subsection{Bandwidth Prediction}
Bandwidth prediction can assist in scheduling models to prefetch larger blocks or select high-priority blocks. What's more, accurate bandwidth prediction will be crucial to adapt variable user requirements in immersive video streaming. In this subsection, we depict the design details of the bandwidth prediction model (see in Fig. \ref{BPM}). At time slot $t$, we send two packets to the client from the packet selection queue, e.g., packet $X$ and packet $Y$. Where $X_s$ and $Y_s$ are the sending time of packet $X$ and packet $Y$, respectively. Similarly, $X_r$ and $Y_r$ are the responding time of packet $X$ and packet $Y$ (receiving an ack or lost signal), respectively. $X_s - t$ is the sending latency at time slot $t$. $Y_r - Y_s$ is the round-trip time (RTT) of packet $Y$. Based on the sending latency, we can accurately calculate network bandwidth at time slot $t$:
\begin{equation}
Thr_t = \frac{A_{X}}{X_s - t}
\end{equation}
where $A_{X}$ indicates the size of packet $X$. Since the sending latency of each packet is very small, it is extremely susceptible to external factors such as network equipment and environmental changes, resulting in large fluctuations in network bandwidth forecasts. So we use a moving average method to predict the network bandwidth as follows:
\begin{equation}
\overline{Thr}_{Y_r} =   \frac{\sum_{i=1}^{Inflight_{(Y_s,Y_r)}}A_i}{Y_r - Y_s}
\end{equation}
where $Inflight_{(Y_s,Y_r)}$ indicates the number of packets that have been sent after time $Y_s$ but have not been received at time $Y_r$. $A_i$ indicates each packet size. When the transmission network is not congested, the server will continue to send packets to the client. $\frac{\sum_{i=1}^{Inflight_{(Y_s,Y_r)}}P_i}{Y_r - Y_s}$ represents the total data size by the server sending over time $[Y_r, Y_s]$. Through experiments, it is verified that it can accurately reflect the changing trend of network bandwidth and reduce the bandwidth oscillation frequency.

\begin{figure*}[t]
	\centering 
	\subfigure[Selecting blocks in sequence.]{
		\centering
		\includegraphics[width=0.32\linewidth]{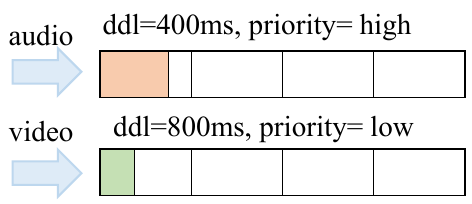}
	}%
	\subfigure[Prefetching the larger data blocks.]{
		\centering
		\includegraphics[width=0.32\linewidth]{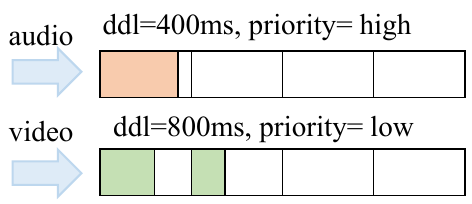}
	}%
	\subfigure[Sending high priority data blocks first.]{
		\centering
		\includegraphics[width=0.32\linewidth]{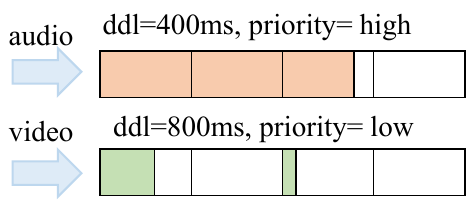}
	}%
	\centering
	\caption{Traffic scheduling strategy.}
	\label{TSS}
\end{figure*}

\subsection{Traffic Scheduling}
In each time slot $t$, the traffic scheduling model can select one block from the block awaiting queue and push the next packet of this block into the packet selection queue. 

We usually select blocks in order. In the high network bandwidth, if we can prefetch larger data blocks in the block awaiting queue (see in Fig. \ref{TSS}), which can make more data blocks meet users' deadline requirements. On the contrary, if we can send higher-priority data blocks in the low network bandwidth, which can alleviate the drastic fluctuations in user experience caused by network bandwidth oscillations. Since the network bandwidth can be accurately predicted by the network bandwidth prediction model, the challenge is how to achieve a higher delivery ratio.

We assume that $Sr_t^i$ is the remaining size ratio of block $Q_t^i$, $Dr_t^i$ is the remaining deadline ratio of block $Q_t^i$, $Pkt_t^i$ is the remaining packet number of block $Q_t^i$, and $P_t^i$ is the priority of block $Q_t^i$. As shown in Algorithm \ref{algo1}, in the block awaiting queue, we first filter out blocks that have exceeded users' deadline. Then, we send retransmission packets first in the block awaiting queue. When the network bandwidth is below a certain threshold $L$, we first send blocks with a larger remaining average deadlines $Dr_t^i / Pkt_t^i$ and higher priority $P_t^i$. When the network bandwidth exceeds a certain threshold $H$, we first send blocks with a larger remaining size ratio $Sr_t^i$, smaller remaining deadline ratio $Dr_t^i$ and lower priority $P_t^i$. In other cases, we first send blocks with smaller remaining size $Sr_t^i*Pkt_t^i$ and blocks with smaller remaining average deadline $Dr_t^i / Pkt_t^i$.

\begin{algorithm}[t]
	\SetKwData{thr}{$\overline{Thr}$}
	\SetKwFunction{Prediction}{Prediction} 
	\SetKwInOut{Input}{input}
	\SetKwInOut{Output}{output}
	\caption{Traffic scheduling scheme}
	\label{algo1} 
	
	\Input{Time slot, $t$;\\
		Block awaiting queue, $Q_t$;\\
		The length of block awaiting queue, $N_t$;} 
	\Output{Best block ID, $B_t$;} 
	\For{$i\leftarrow 0$ \KwTo $N_t$}
	{
		\thr = \Prediction(Inflight, packet{\_}size, rtt)\;
		\tcp*[h]{Special treatment for the first block}\;
		\lIf{$B_t = -1$}{$B_t \Leftarrow i$}
		\lIf{$F_{k,v} - E_{k,v} > D_{k,v}$}{$B_t \Leftarrow B_t$;continue}
		\lIf{$Q_{t,i}$ is a retrans packet}{$B_t \Leftarrow i$;continue}
		\If{\thr belows a certain threshold $L$}
		{\lIf{$Dr^{B_t} / Pkt^{B_t}*(1+P^{B_t}) \le Dr_t^i / Pkt_t^i*(1+P_t^i)$}{$B_t \Leftarrow i$;continue}
			\lElse{$B_t \Leftarrow B_t$;continue}}
		\If{\thr exceeds a certain threshold $H$}
		{\lIf{$Dr^{B_t}*(1+P^{B_t})/Sr^{B_t} \ge Dr_t^i*(1+P_t^i)/Sr_t^i$}{$B_t \Leftarrow i$;continue}}
		\lIf{$Sr^{B_t}*Pkt^{B_t} \le Sr_t^i*Pkt_t^i$}{$B_t \Leftarrow B_t$;continue}
		\lIf{$Dr^{B_t} / Pkt^{B_t} \le Dr_t^i / Pkt_t^i$}{$B_t \Leftarrow B_t$;continue}
		\lElse{$B_t \Leftarrow B_t$}
	}
\end{algorithm}

\section{Experiments}
\subsection{Experimental Setup}\label{methodology}
We collected video traces of three application scenarios: VR battle gaming, VR live streaming, and VR chess gaming. Each video trace contains multiple media elements and different user requirements, as shown in TABLE \ref{video datasets}. The total length of each video trace is more than 20 seconds. 

We used three real network traces datasets: the FCC broadband dataset provided by the US Federal Communications Commission\footnote{https://www.fcc.gov/reports-research/reports/.}, a 3G/HSDPA mobile dataset collected from Norway\cite{3G/HSDPA}, and a 4G/LTE mobile dataset collected from Belgium\cite{4G/LTE}. We created a corpus of 2000 network traces of 500 seconds each by combining three public datasets. To avoid trivial cases for which choosing the maximum sending rate is always the optimal solution or where the network cannot support any available sending rate, we only considered original traces whose average throughput was less than 3 Mbps and whose minimum throughput exceeded 0.2 Mbps. We used a random sample of 80{\%} of our corpus as a training corpus and the remaining 20{\%} as a test corpus. We used cross-validation to verify our scheme's performance.	

We adopt a immersive video streaming transmission scheduling simulation platform\footnote{Simulation platform will be open when this paper is accepted.}. It stimulates the interaction between the viewer and the streaming media server. The client and the streaming media server use Mahimahi's network emulation tools\footnote{http://mahimahi.mit.edu/} to emulate many different link conditions (different round-trip times and random loss packet rate) by taking the real network trace as input.

Based on the evaluation indicators of cooperative enterprise in the video streaming scenarios and the existing transmission scheduling models\cite{frame2, sche, cc1}, the optimal hyperparameters of QoE can be designed as follows: $\alpha=0.85$. Through experimental comparison, our algorithm works best when $H=2.3$ and $L=0.8$. We present performance comparisons between our scheme and three baseline models, including block-based SFRA\cite{BS}, greedy-based LDF\cite{GS}, and buffer-based RSWN\cite{RS2}.

\begin{table}[t]
	\caption{Video datasets.}
	\label{video datasets}
	\centering
	\begin{tabular}{c|c|c|c}
		\toprule
		Scenarios & Elements & Priority & Deadline\\
		\midrule
		Scenario{\_}1&3&[0,1,2]&[0.1,0.15,0.2]\\
		Scenario{\_}2&2&[0,1]&[0.1,0.2]\\
		Scenario{\_}3&3&[0,1,2]&[0.15,0.5,0.2]\\
		\bottomrule
	\end{tabular}
\end{table}

\subsection{Our scheme vs. Existing models}
In this subsection, to verify that our scheme can adapt to user-personalized requirements and improve user satisfaction, we compare it with existing scheduling models in terms of QoE metrics. For comparison, we trained all scheduling models in the simulation platform with BBR\cite{BBR} congestion control algorithm. In each experiment, we continued to adjust the scheduling models until we obtained the optimal model under the QoE metrics considered. The results were collected on the 3G/HSDPA and 4G/LTE hybrid network datasets.

From Fig. \ref{plot_1_1} and Fig. \ref{plot_1_3}, we can determine that our scheme's delivery ratio and QoE are far superior to other existing scheduling models. Because we build an accurate bandwidth prediction model and a more fine-grained transmission scheduling strategy. The accuracy bandwidth prediction model can assist scheduling models to first prefetch larger blocks or select high-priority blocks. In the high network bandwidth, we can prefetch the larger data blocks in the block awaiting queue instead of transmitting them in order, to meet the deadline requirements of more data blocks. On the contrary, we can send higher-priority data blocks to alleviate the drastic fluctuations in user experience caused by network bandwidth oscillations and improve the delivery ratio. From Fig. \ref{plot_1_2}, we can find that our scheme's channel utilization is the same as other existing scheduling models. This is because the existing models contain many invalid block transmissions, which do not transmit the entire block before the deadline. When considering effective channel utilization, our scheme still has a huge advantage.

\subsection{Generalization}
\begin{figure}[t]
	\centering 
	\subfigure[Delivery ratio]{
		\centering
		\includegraphics[width=0.49\linewidth]{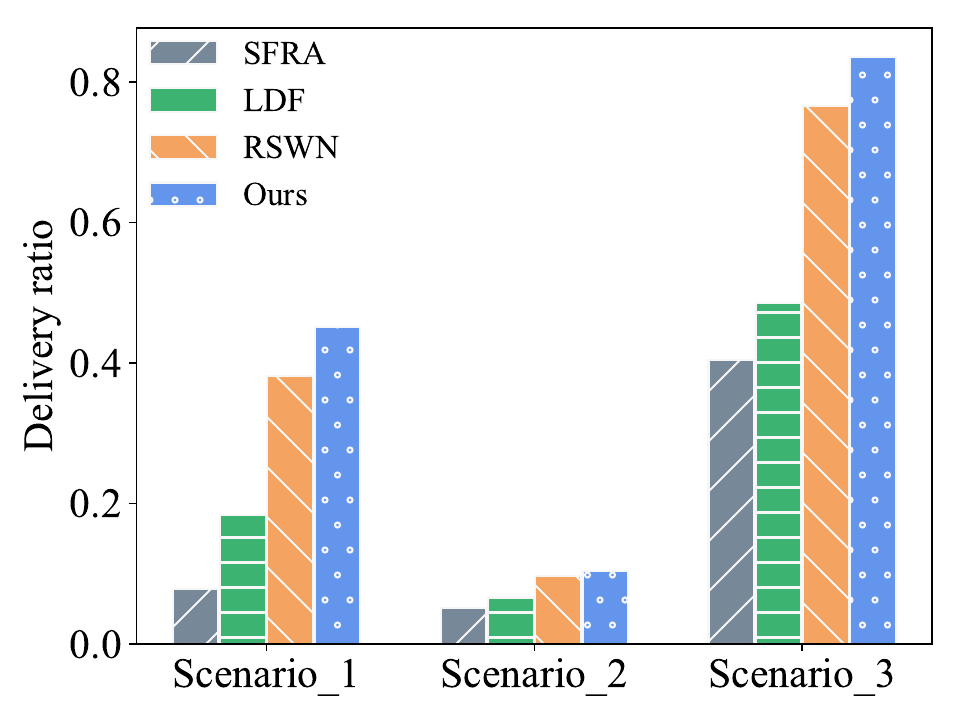}
		\label{plot_1_1}
	}%
	\subfigure[Channel utilization]{
		\centering
		\includegraphics[width=0.49\linewidth]{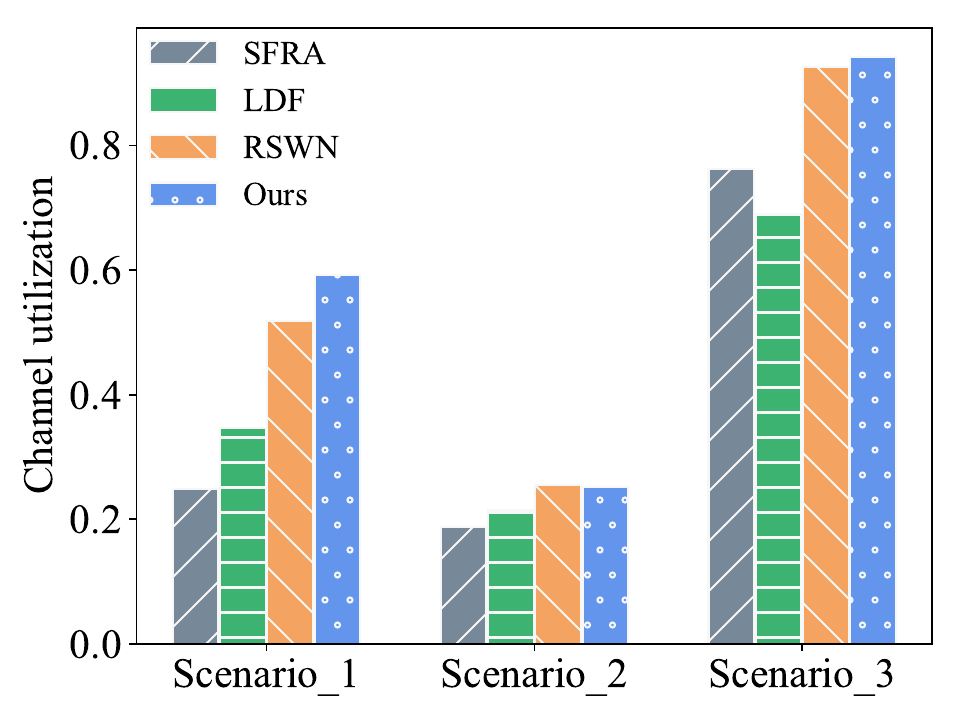}
		\label{plot_1_2}
	}%
	
	\subfigure[QoE]{
		\centering
		\includegraphics[width=0.49\linewidth]{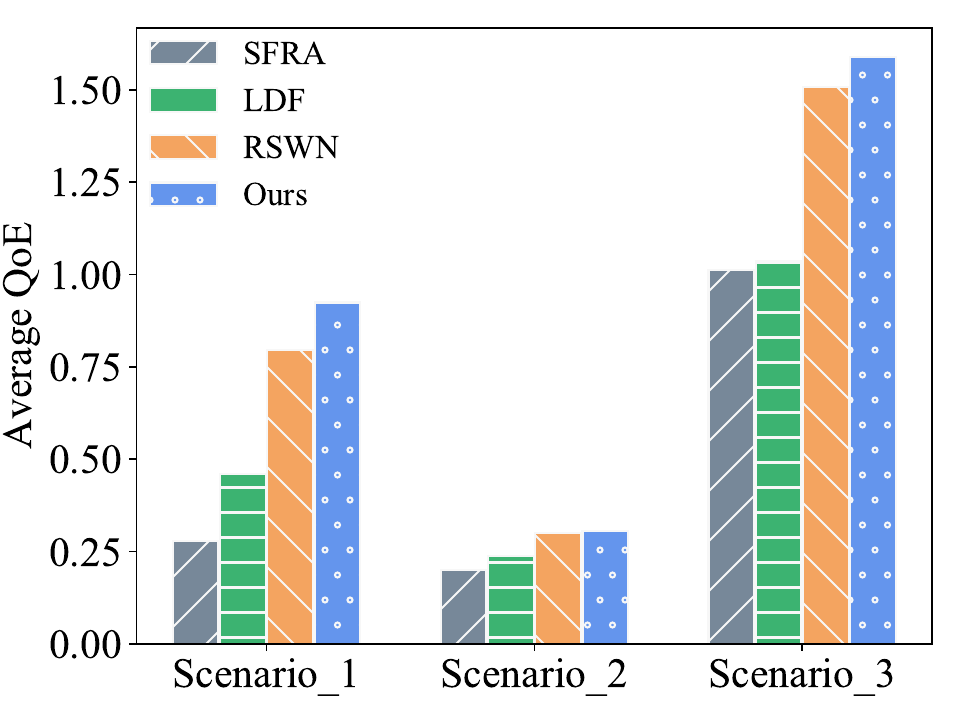}
		\label{plot_1_3}
	}%
	\subfigure[Scenario{\_}1]{
		\centering
		\includegraphics[width=0.49\linewidth]{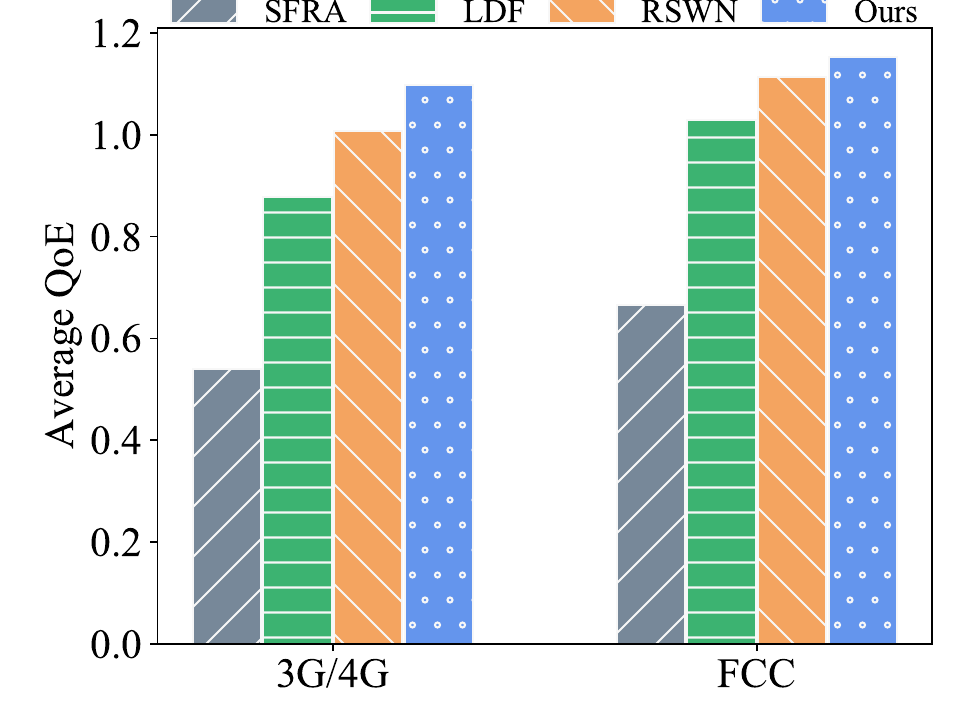}
		\label{plot_1_4}
	}%
	
	\subfigure[Scenario{\_}2]{
		\centering
		\includegraphics[width=0.49\linewidth]{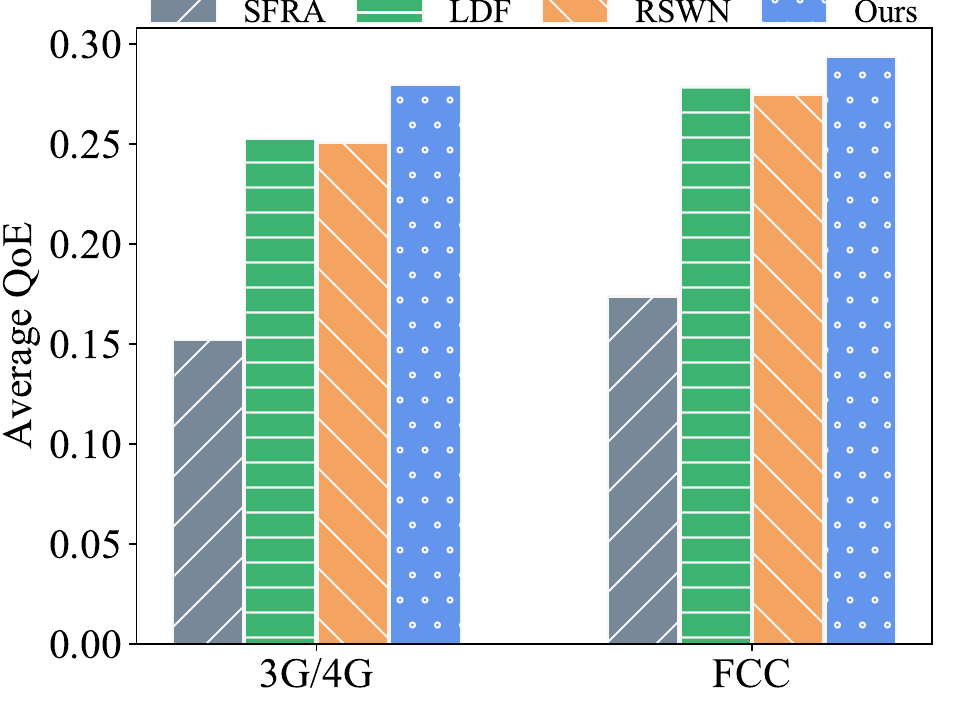}
		\label{plot_1_5}
	}%
	\subfigure[Scenario{\_}3]{
		\centering
		\includegraphics[width=0.49\linewidth]{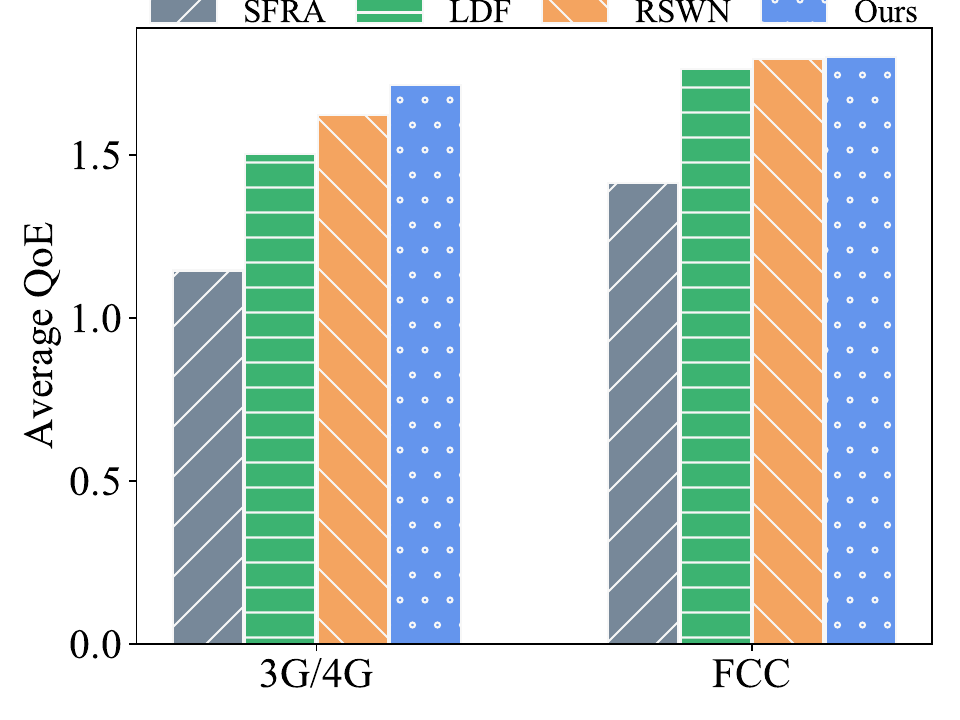}
		\label{plot_1_6}
	}%
	\centering
	\caption{(a-c):Performance comparison between our scheme and existing scheduling models in 3G/HSDPA and 4G/LTE hybrid network datasets. (d-f): Performance comparison between our scheme and existing scheduling models in different video and network scenarios in terms of average QoE.}
	\label{plot_1}
\end{figure}

In this subsection, to verify that our scheme has strong generalizability, we compare it with existing scheduling models in the different video and network scenarios in terms of average QoE. In each experiment, we continued to adjust those models until we obtained the optimal model for the average QoE. For comparison, we trained all scheduling models in the simulation platform with the BBR congestion control algorithm. The experimental results are shown in Fig. \ref{plot_1}.

As shown in Fig. \ref{plot_1_4}, Fig. \ref{plot_1_5} and Fig. \ref{plot_1_6}, our scheme can maintain high performance for each video scenarios and network conditions considered. Especially for the 3G/4G wireless network, it performed significantly better than existing scheduling models with improvements in average QoE of 12{\%} - 31{\%}. This is because our network bandwidth prediction model uses the moving average $\overline{Thr}_t$ of the real network bandwidth $Thr_t$, rather than only building simple and inaccurate models based on cumulative historical bandwidth.

\section{Conclusion}
Deadline-aware transmission scheduling in immersive video streaming is crucial. In this paper, we propose a deadline and priority-constrained immersive video streaming transmission scheduling scheme. It builds an accurate bandwidth prediction model and performs scheduling based on the user's personalized latency sensitivity thresholds and the media element's priority. We evaluate our scheme via trace-driven simulations. In the future, we hope that our scheme can be deployed in the real network system.

\bibliographystyle{./IEEEtran}
\balance
\bibliography{IEEEabrv, DPTS}
\end{document}